\documentclass[12pt]{article}
\usepackage{amsmath}
\input{BoxedEPS}

\SetRokickiEPSFSpecial  
\HideDisplacementBoxes

\numberwithin{equation}{section}
\begin{document}

\title{The Casimir Effect for Fermions in One Dimension}
\author{P.~Sundberg\footnote{Email: sundberg@mit.edu}\enspace
and R.L.~Jaffe\footnote{Email: jaffe@lns.mit.edu} \\
\\
{\small\itshape Center for Theoretical Physics}
  \\[-1ex]
{\small\itshape Department of Physics and
Laboratory for Nuclear Science}\\[-1ex]
{\small\itshape Massachusetts Institute of
Technology} \\[-1ex]
{\small\itshape  Cambridge, Massachusetts 02139} }
\date{\small MIT-CTP-3406}
\maketitle

\pagestyle{myheadings} \markboth{P.~Sundberg and R.L.~Jaffe}{The
Casimir Effect for Fermions in One Dimension}
\thispagestyle{empty}

\begin{abstract}
\noindent

We study the Casimir problem for a fermion coupled to a static
background field in one space dimension.  We examine the relationship
between interactions and boundary conditions for the Dirac field.  In
the limit that the background becomes concentrated at a point (a
``Dirac spike'') and couples strongly, it implements a confining
boundary condition.  We compute the Casimir energy for a masslike
background and show that it is finite for a stepwise continuous
background field.  However the total Casimir energy diverges for the
Dirac spike.  The divergence cannot be removed by standard
renormalization methods.  We compute the Casimir energy density of
configurations where the background field consists of one or two sharp
spikes and show that the energy density is finite except at the
spikes.  Finally we define and compute an interaction energy density
and the force between two Dirac spikes as a function of the strength
and separation of the spikes.

\end{abstract}

\section{Introduction}

When a quantum field is coupled to matter, the 
spectrum of its quantum fluctuations changes. The resulting energies,
forces, and pressures are known in general as Casimir effects and
have a variety of experimental consequences.\cite{casimir, BMM}
In many applications the details of the coupling to materials can
be idealized by boundary conditions, and the results depend on
the geometry and the boundary condition alone.  The Casimir force
between grounded metal plates has now been measured quite
accurately and agrees with Casimir's original
prediction~\cite{BMM}.

In Refs.~\cite{letter,longpaper} new methods were developed to
examine Casimir effects without resorting to boundary conditions
{\it ab initio\/}.  Instead the fluctuating fields are coupled to
a non-dynamical background field, the renormalized effective
energy is calculated using standard methods of quantum field
theory and renormalization theory, and examined in the limit that
the background becomes sharp and strong enough to enforce a
boundary condition on the fluctuating
fields \cite{review}.  In general the renormalized
effective energy diverges in this limit, indicating that the
physical situation depends in detail on the coupling between the
field and the matter.  In the examples studied in
Refs.~\cite{letter,longpaper} forces between rigid objects and
energy densities away from boundaries remain finite and agree
with the boundary condition idealization.  However the Casimir
pressure on an isolated surface was found to diverge in the
boundary condition limit, signalling that it depends in detail on
the properties of the material that provide the physical
ultraviolet cutoff.\cite{newpaper}

In this paper we apply the methods of
Refs.~\cite{letter,longpaper} to fluctuating Dirac fermions in
one dimension.  Fermionic Casimir effects are interesting in
their own right, for example in the bag model of
hadrons.\cite{CJJTW,Milton} Boundary conditions are in general
more disruptive to Dirac fields than boson fields because the
equations of motion are first order, so we anticipate that
fermion Casimir effects will provide an instructive laboratory to
study the difference between the approach of
Refs.~\cite{letter,longpaper} and the boundary condition method.

Our object is to treat the boundary condition as the limiting
form of the coupling to a smooth, finite background potential,
${\cal L}_{\rm int}=\lambda\bar\psi V(x)\psi$.  We
concentrate on potentials, $V(x)$, that couple like the
mass, because such backgrounds lead to a confining (bag) boundary
condition when they become strong.\cite{CJJTW} As in the scalar
case, the boundary condition limit is achieved by letting
$V(x)$ become concentrated at one point (the ``sharp''
limit) -- we refer to this standard configuration as a ``Dirac
spike'' -- and then letting $\lambda\to\infty$ (the ``strong''
limit).  In the Dirac case the renormalized vacuum fluctuation
energy diverges already in the sharp limit.  In the boson case it
was finite for a sharp background in one dimension and diverged
only in the limit $\lambda\to\infty$.  Our results on the energy
density and the force between ``Dirac spikes'' are more positive.
We find that the renormalized vacuum fluctuation energy density
away from the spikes is finite even in the strong limit, and that
the force between Dirac spikes is also finite.

In Section \ref{sc:dirac} we discuss the behavior of the Dirac
equation in the presence of an interaction concentrated at a
single point.  We show that it is most convenient to represent the
interaction by a transfer matrix which relates the Dirac
wavefunction on the immediate left of a singularity to the
wavefunction on the immediate right.  In this way we are able to
define the Dirac spike which is the subject of much of the rest
of our work and which implements a bag boundary condition in the
limit $\lambda\to\infty$.

In Section~\ref{sc:phase shifts} we briefly review the method
developed in Ref.~\cite{firstpaper} to compute the total renormalized
Casimir energy in a background field.  We emphasize that the
divergences that appear in the case of a smooth background are
cancelled by allowed counterterms already present in the
continuum Lagrangian.  We then apply this formalism to compute
the energy associated with a single Dirac spike, and find that it
is ill defined.

In Section~\ref{sc:energydensity} we consider energy densities
and forces between Dirac spikes. First we introduce a method
based on the Greens function\cite{longpaper}, which allows
us to compute the energy density outside the region of
interaction.  We then apply this method to compute
the energy densities of configurations with one Dirac spike at
the origin, and two spikes separated by a distance $d$, both of
which we find to be finite even as $\lambda\to\infty$.  We then
introduce an interaction energy density, defined as the energy
density of the two spike configuration minus twice the energy
density of a single spike, placed to cancel the local divergences
in the energy density, and conclude by integrating the
interaction energy density to compute the total energy.  From
that we derive the force between the two Dirac spikes in one
dimension.

The problem of vacuum fluctuations of the Dirac field in one
dimension has been studied
previously\cite{Johnson,Milton:1983wy}.  In
Ref.~\cite{Bordag,M&T} boundary conditions were imposed
{\it ab initio\/}, and zeta function regularization was used
to control divergences.  In Ref.~\cite{Elizalde:2002wg}, zeta
function regularization was used to derive an expression for the
total energy of a configuration of two parallel plates, also
using boundary conditions.  In~\cite{DePaola:1999im}, expressions
for zero-point energies and energy densities were developed using
a zeta function regularization framework, but the physical
potential implementing the boundary conditions was not taken into
account.  In this paper we implement the boundary conditions as 
the limit of a finite potential.  In  this way we can control the 
subtleties of the calculation at every step enabling us to 
confidently distinguish what is finite and cutoff independent and 
what is divergent in the boundary condition limit.

This work is related to Ref.~\cite{fermions}, which applied the methods of
Ref.~\cite{firstpaper} to fermion fluctuations in a smooth
background in one dimension, but did not consider either the
energy density or the limit of sharp and strong backgrounds
necessary to implement boundary conditions.  In this paper,
we will adopt the renormalization conventions of Ref.~\cite{fermions}
and refer the reader there for further discussion.

\section{The Dirac Equation in a Singular Background}
\label{sc:dirac}

The equation of motion for the fluctuations of a Dirac field in a
non-dynamical background field, $V$, in one dimension
reads
\begin{equation}
\label{eq:dirac}
\left[-i\alpha \frac{d}{dx}+\beta m + V(x)\right] \psi(x)
= \omega \psi(x)
\end{equation}
Here, $\alpha = \gamma_0 \gamma_1$, $\beta = \gamma_0$ are
two-by-two matrices, and $\omega$ is the energy eigenvalue of the
time-independent solution $\psi(x)$ given the potential
$V(x)$.  We work in a chiral basis, $\alpha = \sigma_3$,
$\beta = \sigma_1$, $\gamma_5 = \gamma_0\gamma_1 = \sigma_3$.
Our spinor normalization for right- ($k>0$) and leftgoing ($k<0$)
waves (when $V=0$) is determined by 
\begin{equation}
\label{eq:dirac-free-soln}
        \psi_{0}(k,x)\equiv v(k)e^{ikx}=
        \begin{pmatrix}
                \sqrt{\omega+k} \cr
                \sqrt{\omega-k}
        \end{pmatrix} e^{ikx}
\end{equation}
with $k=\sqrt{\omega^2-m^2}$.  Note that $V(x)$ is
matrix-valued.  In its most general form,
\begin{equation}
        V(x) = V_0(x)I + V_1(x)\alpha +
        V_2(x)\beta + V_3(x)\alpha\beta\, ,
\end{equation}
where $V_{0}$, $V_{1}$, and $V_{2}$ are real
and $V_{3}$ is imaginary. For
simplicity we exclude backgrounds proportional to $\alpha\beta$.
This eliminates interactions of the form ${\cal L}_{\rm
int}=\bar{\psi}\gamma_5V_{3}(x)\psi$.  In addition,
$V_1$ can  be gauged away.  To show this,
assume that $\psi(x)$ solves eq.~(\ref{eq:dirac}) when
$V_1(x)=0$.  Then it is easy to show that
\begin{equation}
        \chi(x) = A(x)\psi(x) = e^{-i\int_0^x
        V_1(x')dx'}\psi(x)
\end{equation}
solves eq.~(\ref{eq:dirac}) for any $V_1(x)$.  Since
$V_1$ only affects the phase of the solutions (and is
independent of energy), the density of states as a function of
$x$ and $\omega$ is independent of $V_1$, and we conclude
that the energy density and hence the total energy are
independent of $V_1$.  We can therefore, without further
loss of generality, consider only potentials that can be written
as a linear combination of the identity and $\beta$.  To
demonstrate the physical significance of the two remaining terms,
reorder the terms in eq.(\ref{eq:dirac}),
\begin{equation}
\left[-i\alpha \frac{d}{dx}+\beta (m+M(x))\right] \psi(x) =
 (\omega-\phi(x))\psi(x).
\end{equation}
where we have replaced $V_{0}(x)$ by $\phi(x)$ and
$V_{2}(x)$ by $M(x)$

The component proportional to $\beta$ in effect acts like an
extra mass term and the component proportional to the identity
acts like an electrostatic potential.  The mass term affects
positive and negative energy eigenstates identically, both are
attracted or both repelled.  The electrostatic potential, on the
other hand, treats positive and negative energy eigenstates in an
opposite manner, attracting one and repelling the other.  A
significant portion of the later discussion will be focused on
the pure mass-like potential.

As explained in the Introduction, we intend to implement a
boundary condition as the limit of a finite potential.  We must
first investigate how this can be done for the Dirac equation in
a consistent way. For
simplicity, whenever the explicit form of the finite potential is
needed, we will use a rectangular barrier.  As the barrier width
$a$ goes to zero while its area remains constant, the potential
can be replaced by a transfer matrix relating $\psi$ at the left
of the barrier to $\psi$ at the right.  We call this object a
``Dirac spike''.  If we then let the area under the barrier go to
infinity, the domains to the left and right of the spike decouple
and $\psi$ obeys a flux conserving boundary condition at the
spike.  Depending on the matrix nature of the potential, the
Dirac spike may have up to two bound states.

In the scalar case the representation of a pointlike potential
is simple: a delta
function potential serves the purpose.  Since the Dirac equation
is first order, however, a delta function potential leads to
contradictions.  For example, consider a masslike potential
$V(x)=\beta M\delta(x)$.  Written out in components the Dirac
equation becomes
\begin{eqnarray}
        \label{eq:inconsistent-1}
        -i \frac{d\psi_1}{dx} + (m+M\delta(x))\psi_2 = \omega\psi_1 \\
        \label{eq:inconsistent-2}
        i \frac{d\psi_2}{dx} + (m+M\delta(x))\psi_1 = \omega\psi_2.
\end{eqnarray}
The terms involving a delta-function are only well defined if
$\psi$ is continuous at $x=0$.  However the first equation
implies a jump in $\psi_1$ for continuous $\psi_2$, and the
second requires a jump in $\psi_2$ for continuous $\psi_1$.  Thus
the equations are not consistent.

Anticipating that a sharp potential will generate a discontinuity
in $\psi$ at the point of interaction, we study a sharp potential
as the limit of a finite one.  Consider a potential of the form
\begin{equation}
\label{eq:box-potential}
V(x,a)=\Gamma b(x,a),
\end{equation}
where $b(x,a)$ is defined to be $1/a$ for $0 \le x \le a$ and $0$
otherwise, and $\Gamma$ is some matrix-valued constant.  It
is easy to show that the solution to the Dirac equation in a
potential of the form of eq.~(\ref{eq:box-potential}) is a
continuous function of $x$.

Since the Dirac equation is first order, $\psi(x)$ is completely
determined from its value at any point.  In particular, given
$\psi(0)$, we can compute $\psi(a)$, $\psi(a)= T(\Gamma,a,\omega)
\psi(0)$, where the two dimensional transfer matrix, $T$, depends
on $\omega$, $a$, and $\Gamma$.  As long as $a$ is finite, $\psi$
is continuous everywhere.  But as we take the limit $a \to 0$,
keeping $\Gamma$ fixed, $\psi$ is no longer continuous at $x=0$.
We therefore define the transfer matrix for a Dirac spike by
$T(\Gamma,\omega)=\lim_{a \to 0}T(\Gamma, a, \omega)$ with the
jump condition
\begin{equation}
\label{eq:jump-condition}
        \psi(0^+)=T(\Gamma,\omega)\psi(0^-).
\end{equation}
where $\psi(0^{\pm})=\lim_{x\to 0^{\pm}}\psi(x)$.

The properties of $T$ depend on both the strength and Dirac matrix
character of $\Gamma$.  As advertised we limit our discussion to
potentials of the form,
\begin{equation}
        \Gamma=\theta I+ \lambda\beta,
\end{equation}
where $\theta=\phi a$ and $\lambda=Ma$ are constants.
For finite $a$ we obtain
\begin{equation}
\label{eq:transfer-matrix}
    T(M,\phi,a,\omega)
    =\begin{pmatrix}\cos qa +i\frac{\omega-\phi}{q}\sin qa &
    -i\frac{m+M}{q}\sin qa \cr
    i\frac{m+M}{q}\sin qa &
    \cos qa -i\frac{\omega-\phi}{q}\sin qa\end{pmatrix}
\end{equation}
where $q = \sqrt{(\omega-\phi)^2 - (m+M)^2}$.  In the limit $a\to
0$ \emph{ for any fixed value of $\omega$}, we obtain the
transfer matrix for the general Dirac spike,
\begin{equation}
\label{eq:full-transfer-matrix}
    T(\lambda,\theta) =\begin{pmatrix}
    \cos\sqrt{\theta^2-\lambda^2}-i\frac{\theta}{\sqrt{\theta^2-\lambda^2}}
    \sin\sqrt{\theta^2-\lambda^2} &
    -i\frac{\lambda}{\sqrt{\theta^2-\lambda^2}}\sin\sqrt{\theta^2-\lambda^2}\cr
    i\frac{\lambda}{\sqrt{\theta^2-\lambda^2}}\sin\sqrt{\theta^2-\lambda^2} &
    \cos\sqrt{\theta^2-\lambda^2}+i\frac{\theta}{\sqrt{\theta^2-\lambda^2}}
    \sin\sqrt{\theta^2-\lambda^2}
    \end{pmatrix}
\end{equation}
This is the generalization of the delta function potential to the
Dirac equation (modulo our exclusion of pseudoscalar couplings).
Note that in the limit $a\to 0$ $T$ becomes independent of
$\omega$ and can therefore be regarded as a background
``potential'' in the Dirac equation.

Two special cases are of interest.  The masslike case, where
$\theta=0$, will play the principal part in the sections that
follow,
\begin{equation}
\label{eq:transfer-matrix-exp}
    T_{\rm M}(\lambda)
    =\begin{pmatrix}\cosh \lambda & -i \sinh \lambda \cr
    i \sinh \lambda & \cosh \lambda \end{pmatrix} =
    e^{i \gamma_1 \lambda}
\end{equation}
For completeness, we will also compute the electrostatic
case, where $\lambda=0$,
\begin{equation}
\label{eq:transfer-matrix-exp-phi}
T_{\rm E}(\theta)
    =\begin{pmatrix}e^{-i\theta} & 0 \cr
    0 & e^{i\theta}\end{pmatrix} =
    e^{-i \gamma_5 \theta}
\end{equation}

These two potentials may or may not have bound states.  For the
pure mass-like case, there are positive and negative energy bound
states when $\lambda<0$.  The energy eigenvalue obeys
\begin{equation}
\label{eq:bound-state-lambda}
        \frac{\sqrt{m^2-\omega_{\rm M}^2}}{m}\equiv
        \frac{\kappa_{M}}{m}=-\tanh\lambda,
\end{equation}
whereas in the electrostatic case, there is one bound state
in either the positive or negative energy spectrum regardless of
the sign of $\theta$,
\begin{equation}
\label{eq:bound-state-theta}
        \frac{\sqrt{m^{2}-\omega_{\rm E}^{2}}}{\omega_{\rm E}}\equiv
        \frac{\kappa_{\rm E}}{\omega_{\rm E}}=-\tan \theta
\end{equation}

Equations~(\ref{eq:bound-state-lambda})
and~(\ref{eq:bound-state-theta}) are qualitatively different.
Whatever the sign of the electrostatic interaction, it is
attractive either for particles (positive energy states) or
antiparticles (negative energy states).  The electrostatic
interaction violates charge conjugation invariance so the
spectrum need not be symmetric, however it is periodic in
$\theta\to\theta+2\pi$.  The masslike spike has either no bound
states when it is repulsive, or exactly one positive energy and
one negative energy bound state when it is attractive.  The
energy spectrum is symmetric since a masslike interaction
preserves charge conjugation invariance.

The probability current passing through a masslike spike goes to
zero as the strength of the spike goes to infinity.  Effectively
the spike becomes a wall, splitting the line into two independent
half lines.  This situation is familiar from bag models, and not
surprisingly the masslike spike reduces to a bag boundary
condition in this limit.\cite{CJJTW} In contrast the
electrostatic spike is periodic in the strength $\theta$ and
therefore does not reach a limit as $\theta\to\infty$.  While
this phenomenon is interesting in its own right, we do not pursue
it any further here.

As $\lambda \to \infty$, eq.~(\ref{eq:transfer-matrix-exp})
becomes
\begin{equation}
        T_{\rm M}(\lambda)=e^\lambda
        \begin{pmatrix}
                1 & -i \cr -i & 1
        \end{pmatrix}+{\cal O}(e^{-\lambda})
\end{equation}
For $\psi(0^+)$ to be finite, it is necessary that

\begin{equation}
\label{eq:boundary-}
        \psi_1(0^-)=i\psi_2(0^-)+{\cal O}(e^{-\lambda}),
\end{equation}
from which it follows that
\begin{equation}
\label{eq:boundary+}
\psi_1(0^+)=-i\psi_2(0^+)+{\cal O}(e^{-\lambda}).
\end{equation}
Eqs.~(\ref{eq:boundary-}) and~(\ref{eq:boundary+}) can be
written as two independent boundary conditions on the half lines
separated by a true boundary at $x=0$,
\begin{eqnarray}
        (1+i\gamma_{1})\psi(0^+) & = &0\\
        (1-i\gamma_{1})\psi(0^-) & = &0\, ,
\end{eqnarray}
which are standard bag boundary conditions and insure that the
probability current, $j=\bar{\psi}\gamma_1\psi$ vanishes at $x=0$.

\section{Total Renormalized Vacuum Fluctuation Energy}
\label{sc:phase shifts}

In this section we briefly review the method developed in
Ref.~\cite{firstpaper,letter} to compute the Casimir energy of a
field configuration from scattering data, and adapt it to the
case of a Dirac fermion in a static background in 1+1 dimensions.
Our work is based on the prescription of Ref.~\cite{fermions},
specialized to a sharply peaked background.  We first review the
fundamentals of the method, including how the phase shifts are
related to the density of states.  We then show how the resulting
highly oscillating energy integrals can be evaluated using
contour integration, significantly simplifying the numerical
computations.

We first confront the case of a Dirac spike directly and find
that it generates a divergent renormalized vacuum fluctuation
energy.  We then compute the total energy of a potential of the
form of eq.~(\ref{eq:box-potential}), only afterwards taking the
potential to be sharp and strong.  Using this approach, the total
energy is well defined for finite width, $a$, of the potential,
but diverges as $a \to 0$ with the integral of the potential
fixed.

In one dimension the $S$ matrix can be written in terms of the
transmission and reflection coefficients,
\begin{equation}
\label{eq:s-TR}
S=\begin{pmatrix}T & R_1 \cr R_2 & T\end{pmatrix}.
\end{equation}
The transmission coefficients for right and leftgoing waves must
be identical by time reversal symmetry, but the two reflection
coefficients can differ.  The $S$ matrix is constrained to be
unitary, $S^{\dagger}S=I$.  In addition, as $k$ becomes large,
any finite potential becomes negligible, and we expect the $S$
matrix to become the identity in this limit.  Of course we are
considering singular potentials, so this property must be studied
carefully.  In a basis of eigenchannels, the S matrix is
diagonal, and its eigenvalues are complex exponentials of the
phase shifts,
\begin{equation}
\label{eq:s-exp}
        S=\begin{pmatrix}e^{2i\delta_1(k)} & 0 \cr 0 &
        e^{2i\delta_2(k)}\end{pmatrix}.
\end{equation}
The change in the density of states, $\Delta\rho(k)$, due to the
interaction is related to the trace-log of $S$,
$\Delta(k)=\frac{1}{2i}{\rm Tr}\ln S(k)
=\delta_{1}(k)+\delta_{2}(k)$ \cite{firstpaper},
\begin{equation}
\label{eq:rho-from-S}
        \Delta\rho(k)=\rho(k)-\rho^0(k)=\frac{1}{\pi}
        \frac{d\Delta(k)}{dk}
        =\frac{1}{2\pi i}\frac{d\ln\det S}{dk}.
\end{equation}

Following Refs. \cite{review} and \cite{fermions} we write
the \emph{renormalized} Casimir energy for any non-singular
background, $V(x)$, as
\begin{equation}
        \Delta E_{\rm cas} = -\sum_j (\omega_j-m) - \int_0^\infty dk
        (\omega(k)-m)\Delta\overline\rho(k)+ \overline\Gamma_{\rm FD},
        \label{fullcasimir}
\end{equation}
Note the negative sign and factor of two relative to the Casimir
energy of a real scalar field.  The sum is over possible bound
states, and
\begin{equation}
        \Delta\overline\rho(k)= \frac{1}{\pi}\frac{d}{dk}
        \overline{\Delta}(k)\equiv \frac{1}{\pi}\frac{d}{dk}
		\left(\Delta(k)-\Delta^{(1)}(k)\right),
\end{equation}
where $\Delta^{(1)}(k)$ is the first Born approximation to
$\Delta(k)$.  The subtraction of the first Born approximation is
compensated by adding back the corresponding Feynman diagrams,
which are then combined with the counterterms to give the
renormalized Feynman diagram contribution, $\overline\Gamma_{\rm
FD}$.  We work in the ``no tadpole'' renormalization scheme where
the counterterms exactly cancel the local, divergent Feynman
diagrams.  In this renormalization scheme $\Gamma_{\rm FD}=0$.
For more discussion of the derivation of eq.~(\ref{fullcasimir})
we refer the reader to Ref.~\cite{fermions}.

The formalism of Ref.~\cite{review} and
eq.~(\ref{fullcasimir}) gives a finite renormalized energy for
any finite potential.  However, the total energy of a Dirac spike
is infinite.  This is easy to see by computing the associated
shifted density of states and and examining its behavior at large
$k$.

The $S$ matrix can easily be obtained from the transfer matrix,
eq.~(\ref{eq:transfer-matrix-exp}).  The result is
\begin{equation}
        \label{eq:s-matrix}
        S=\begin{pmatrix}T & R_1 \cr R_2 & T\end{pmatrix} =
        \frac{1}{k \cosh \lambda + i m \sinh \lambda}
        \begin{pmatrix}k & -i\omega \sinh\lambda \cr -i\omega \sinh
        \lambda & k\end{pmatrix}.
\end{equation}
It is already apparent that the Dirac spike is a badly behaved.
We expect $S\to I$ as $k\to \infty$, but instead
$$
        S\to \begin{pmatrix}{\rm sech} \lambda & -i\tanh\lambda\cr
        -i\tanh\lambda & {\rm sech}\lambda \end{pmatrix}.
$$
Furthermore, subtraction of the first Born approximation, {\it
ie.\/} of the terms linear in $\lambda$, does not suffice to
render the $k$ integral convergent.  Specifically
\begin{equation}
\label{eq:phase-shift-1}
        \Delta\rho(k)=\frac{1}{\pi}\frac{d}{dk}\Delta(k)=\frac{1}{\pi}
        \frac{m\sinh\lambda\cosh\lambda}{k^2\cosh^2\lambda +
        m^2\sinh^2\lambda}.
\end{equation}
so the integrand in eq.~(\ref{fullcasimir}),
$(\omega(k)-m)\Delta\overline\rho(k)$, is proportional to
$\frac{\tanh \lambda-\lambda}{k}$ at large $k$, and the $k$
integration diverges logarithmically.  It would be necessary to
subtract a term proportional to $\tanh\lambda$ in order to render
the integral finite, and there is no justification in
conventional field theory for a subtraction that includes all
orders in $\lambda$.  We conclude that the Casimir energy
diverges for a Dirac spike, even at finite $\lambda$.

Having encountered a problem confronting the Dirac spike
directly, we back off and instead compute the total
renormalized Casimir energy of a finite mass barrier, and study
the difficulties that develop as the potential becomes sharp.  We
first compute the shift in the density of states by calculating
$\Delta(k)=\frac{1}{2i}\ln\det S$.  The S matrix for this
configuration is
\begin{equation}
        S = \frac{1}{kq \cos qa + i(mM-k^2)\sin qa}
        \begin{pmatrix}kq~e^{-ika} & -iM\omega\sin qa \cr
        -iM\omega\sin qa~e^{-2ika} & kq~e^{-ika}\end{pmatrix}
\end{equation}
From this we conclude that
\begin{eqnarray}
\label{eq:phase-shift-2}
        \frac{d}{dk}\Delta(k) = \frac{1}{2i}\frac{d}{dk}\ln\det S & =
        & \frac{aM(k^2m + k^2M \cos^2 qa - m^2M\sin^2 qa)} {k^2 q^2 +
        M^2 \omega^2 \sin^2 qa} - \nonumber \\
        & & \frac{M^2\sin qa \cos qa (k^2+mM+2m^2)} {k^2 q^3 + M^2
        \omega^2 q \sin^2 qa}
\end
{eqnarray}
Taking $M \to \infty, a \to 0$, with $aM = \lambda$ fixed, sends
$qa \to i\lambda$, and eq.~(\ref{eq:phase-shift-2})
becomes eq.~(\ref{eq:phase-shift-1}), as it must.  The first Born
approximation is given by\footnote{The first Born approximation in
the Dirac theory is equal to the tadpole plus the local part of the
self-energy Feynman Diagram. The explicit form can be derived by writing
the Dirac equation as a second order equation and applying the general
formalism.\cite{fermions}}
\begin{equation}
\label{eq:born}
        \Delta^{(1)}(k) = -\frac{aM}{k}(m+\frac{M}{2}).
\end{equation}
We compute the Casimir energy from eq.~(\ref{fullcasimir})
\begin{eqnarray}
\label{E-ph}
E_{\rm Cas}(m,M,a) & = &-\frac{1}{\pi}\int_0^\infty (\omega-m)
\frac{d}{dk} \left( \Delta(k)-\Delta^{(1)}(k) \right) dk
\nonumber \\
& = & -\frac{aM^2}{\pi}\int_0^\infty \frac{dk (\omega-m)}{k^4 q^3
+ k^2 q \omega^2 M^2 \sin^2 qa}
\left( \frac{k^4q}{2} \cos 2qa\right.\nonumber\\
& + & k^2q(\frac{M^2}{2}+2mM+2m^2) -  
q \sin^2 qa (\omega^2 M (m+\frac{M}{2}) + m^2
k^2)\nonumber\\
&-&\left. \phantom{\frac{1}{2}}\!\!\!\!
k^2 \sin qa \cos qa (k^2+mM + 2m^2) \right)
\end{eqnarray}
It is easy to see that $E_{\rm Cas}(m,M,a)$ is finite
for finite $M$, as it must be, but it diverges as $\lambda M \ln
\frac{M}{m}$ as $a\to 0$ with $\lambda=Ma$ fixed.

\section{Greens Functions and the Energy Density}
\label{sc:energydensity}

The phase shift formalism does not allow us to compute energy
densities.  In this section we extend the formalism of
Ref.~\cite{longpaper} based on the Greens function to enable us
to study energy densities.  We relate the trace of the Greens
function to the \emph{local} density of states and integrate over
$k$ to obtain the local energy density.  We then construct the
Greens function from solutions to the Dirac equation.  We compute
the free Greens function and the interacting Greens function for
one and two spike potentials, and finally use the result to
compute the force between two Dirac spikes.

In one dimension, the Greens function $S(x,x',\omega)$ is a
$2\times 2$ matrix, and obeys

\begin{equation}
(-\omega-i\alpha\frac{d}{dx} +
V(x))S(x,x',\omega)=\delta(x-x')
\end{equation}
We can write $S$ as
\begin{equation}
\label{eq:fermion-green-ie}
S(x,x',\omega)=\sum_n \frac{\psi_n(x)\psi_n^\dagger(x')}
{\omega_n-\omega-i\epsilon}
\end{equation}
where
\begin{equation}
(-i\alpha\frac{d}{dx} + V(x))\psi_n=\omega_n\psi_n
\end{equation}
and
\begin{equation}
\label{eq:fermion-normalize}
\sum_n\psi_n(x)\psi_n^\dagger(x')=\begin{pmatrix}1 & 0 \cr 0 &
1\end{pmatrix}\delta(x-x').
\end{equation}

From eq.~(\ref{eq:fermion-green-ie}) we define a \emph{local}
density of states,
\begin{equation}
\label{eq:rho-S}
        \rho(x,\omega)\equiv\frac{1}{\pi}\mbox{~Im~Tr~}S(x,x,\omega)
        =\sum_n \delta(\omega_n -
\omega)\psi_n^\dagger(x)\psi_n(x)
\end{equation}
We can use the local density of states to compute the local
energy density (or indeed any other local density functional of
the Dirac field).  To compute the energy density we
integrate the density of states weighted by the energy.  However,
we must remember that the ground state energy for a fermion is
$-\omega$, so
\begin{equation}
\label{eq:e-density-omega}
        \epsilon(x)=-\frac{1}{\pi} \int_0^\infty d\omega~\omega
        \mbox{~Im~Tr~}\overline S(x,x,\omega)=
        -\frac{1}{\pi}\int_0^\infty dk~k
        \mbox{~Im~Tr~}\overline S(x,x,\omega(k))
\end{equation}
where $\overline S(x,x',\omega)\equiv
S(x,x',\omega)-S_{0}(x,x',\omega)$, where $S_{0}$ is the free
Greens function.

The free Greens function $S_0$ is given by
\begin{eqnarray}
\label{eq:green-free}
S_0(x,x',\omega)=\frac{i\omega}{k}\left\{\theta(x'-x)v(-k)v^\dagger(k)
\beta~e^{ik(x'-x)} + \right. \nonumber \\
\left.
\theta(x-x')v(k)v^\dagger(-k)\beta~e^{ik(x-x')}\right\}.
\end{eqnarray}
where $v(k)$ is the free Dirac spinor defined by
eq.~(\ref{eq:dirac-free-soln}).  For a configuration with one
spike, the Greens function is
\begin{equation}
\label{eq:green-one-spike}
        S_{1}(x,x',\omega) =  S_0(x,x',\omega) + \frac{\omega^{2}}{k}
        \frac{\sinh\lambda}{k\cosh\lambda+im\sinh\lambda}
		~v(k)v^\dagger(k)\beta~e^{ik(x+x')}
\end{equation}
and for a configuration with two spikes located at $x=\pm d/2$ it becomes
\begin{eqnarray}
        S_2(x,x',\omega) & = &
        S_0(x,x',\omega) - \frac{i\omega^3}
        {k^3}T_2(\omega)\sinh^2\lambda~e^{2ikd}
        \times \nonumber\\
        & & \left(v(k)v^\dagger(-k)\beta~e^{ik(x-x')}+
        v(-k)v^\dagger(k)\beta~e^{ik(x'-x)}\right) +
        \nonumber \\
        &
        &\frac{\omega^2}{k^2}T_2(\omega)
        (\cosh\lambda+\frac{im}{k}\sinh\lambda)
        \sinh\lambda~e^{ikd}\times\nonumber\\
        & & \left(v(k)v^\dagger(k)\beta~e^{ik(x+x')}+
        v(-k)v^\dagger(-k)\beta~e^{-ik(x+x')}\right),
\end{eqnarray}
between the spikes and
\begin{eqnarray}
		S_2(x,x',\omega) & = & S_0(x,x',\omega)+\frac{\omega^2
		T_2(\omega)}{k^3} v(k)v^\dagger(k)\beta~e^{ik(x+x')}
		\times \nonumber \\
        & & (k \sinh 2\lambda \cos kd + 2m\sinh^2\lambda \sin kd)
\end{eqnarray}
outside the spikes. Here $T_2(\omega)$ is the transmission
coefficient for the entire two-spike potential, given by
\begin{equation}
  T_2(\omega)=\frac{k^2}{(k\cosh\lambda+im\sinh\lambda)^2+
    \omega^2\sinh^2\lambda~e^{2ikd}}
\end{equation}

We combine these expressions for the Greens functions with the
definition of the energy density, eq.~(\ref{eq:e-density-omega})
to obtain the energy density for the one and two Dirac spike
configurations.  Note that no subtractions are needed away from
the spikes.  The energy densities are finite, and anyway, the
subtractions are local in the background field, and therefore
vanish except at the spikes.

For the single spike configuration, it follows from
eq.~(\ref{eq:green-one-spike}) and the symmetry under $x \to -x$
that
\begin{equation}
        \mbox{Tr}\{S_1(x,x,\omega)-S_0(x,x,\omega)\}=
        \frac{m\omega}{k}\frac{\sinh\lambda~e^{2ik|x|}}
        {k\cosh\lambda+im\sinh\lambda}
\end{equation}
for all $x \neq 0$.
Using eq.~(\ref{eq:e-density-omega}), we can now write down an
expression for
the energy density at $x\neq 0$,
\begin{equation}
\label{eq:integrand-osc}
        \epsilon_1(x) = -\frac{1}{2\pi} \int_{-\infty}^\infty
        \mbox{Im} \frac{m\omega\sinh\lambda~e^{2ik|x|}}
        {k\cosh\lambda+im\sinh\lambda} dk
\end{equation}
For $x\ne 0$, the integrand of
eq.~(\ref{eq:integrand-osc}) falls off exponentially in the upper
half $k$-plane, but oscillates rapidly on the real axis.  It is
therefore more readily evaluated using the method of contour
integration.  The integrand has a branch cut on the positive
imaginary axis, but no poles in the upper half plane because
there are no bound states, thus
\begin{figure}
\begin{center}
\BoxedEPSF{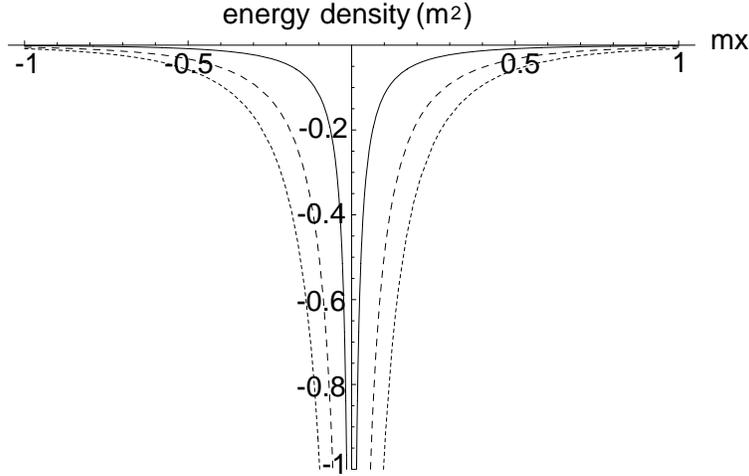 scaled 600}
\end{center}
\caption{Energy density for configurations with a Dirac spike of
strength $\lambda=0.1$ (solid), $\lambda=0.5$ (dashed), and as
$\lambda \to \infty$ (dotted).  The energy density converges
uniformly as $\lambda \to \infty$.}
\label{fig:energy density}
\end{figure}
\begin{equation}
        \label{eq:epsilon1}
        \epsilon_1(\xi) = -\frac{m^2}{\pi}\int_1^\infty
        \frac{\sqrt{\tau^2-1}~\sinh\lambda~e^{-2\tau|\xi|}}
        {(\tau\cosh\lambda+\sinh\lambda)}d\tau
\end{equation}
where $\xi = mx$.  Figure~\ref{fig:energy density} shows a plot
of the energy density for several choices of $\lambda$.
The divergence at $x=0$ is clearly visible for all values of
$\lambda$.

As a check on this method we compute the integrated change in the
density of states $\Delta\rho(k)$ by integrating $\rho(x,\omega)$
over $x$.
\begin{equation}
\Delta \rho(k) =\int_{-\infty}^\infty dx~\rho(x,k) =
\frac{1}{\pi}\int_{-\infty}^\infty \mbox{Im}
\frac{m\sinh\lambda~e^{2ik|x|}}{k\cosh\lambda+im\sinh\lambda} dx.
\end{equation}
The integrand is even in $x$, so we evaluate it only for
positive $x$, and then multiply the result by two.  At the upper
limit, we take $x \to \infty(1+i\epsilon)$, and get no
contribution, yielding the result
\begin{equation}
        \Delta \rho(k) = \frac{1}{\pi}
        \frac{m\omega\sinh\lambda\cosh\lambda}
        {k^2\cosh^2\lambda+m^2\sinh^2\lambda}.
\end{equation}
identical to eq.~(\ref{eq:phase-shift-1}) as expected.

\begin{figure}
\begin{center}
\BoxedEPSF{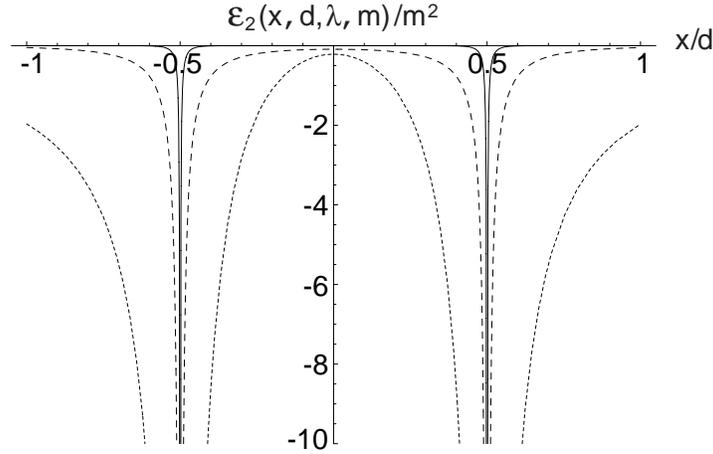 scaled 600}
\caption{Energy density in units of $m^{2}$ for a configuration
of Dirac spikes of strength $\lambda=1$ of separation $d$, with $m=0.1/d$
(dotted), $m=1/d$ (dashed) and $m=10/d$ (solid).}
\label{fig:energy density 2}
\end{center}
\end{figure}

When we turn to two spikes, we maintain parity invariance by
giving the two spikes the same strength parameter $\lambda$, and
placing them at $x=\pm d/2$.  The energy density is even under $x
\to -x$.  Hence, there are only two distinct regions: the region
between the spikes, and the region outside them.  Written as
integrals, the energy densities are
\begin{eqnarray}
        \hbox{For}\quad |x|<d/2,\ \epsilon_2(x,d,\lambda,m) =
        -\frac{1}{\pi}&\int_m^\infty dt 
        \frac{m\sqrt{t^2-m^2}(t\sinh 2\lambda+
        2m\sinh^2\lambda)e^{-td}\cosh 2t|x|}
        {(t\cosh\lambda+m\sinh\lambda)^2+(t^2-m^2)
        \sinh^2\lambda~e^{-2td}}+\nonumber
        \\
        &\frac{2(t^2-m^2)^{3/2}\sinh^2\lambda~e^{-2td}}
        {(t\cosh\lambda+m\sinh\lambda)^2+(t^2-m^2)
        \sinh^2\lambda~e^{-2td}}
        \nonumber\\
        \hbox{For}\quad |x|>d/2,\ \epsilon_2(x,d,\lambda,m) =
        -\frac{1}{\pi}&\int_m^\infty dt 
        \frac{mt\sqrt{t^2-m^2}\sinh 2\lambda \cosh td~e^{-2t|x|}}
        {(t\cosh\lambda+m\sinh\lambda)^2+
        (t^2-m^2)\sinh^2\lambda~e^{-2td}}+\nonumber\\
        &\frac{2m^{2}\sqrt{t^2-m^2}\sinh^2\lambda\sinh td~
        e^{-2t|x|}}{(t\cosh\lambda+m\sinh\lambda)^2+
        (t^2-m^2)\sinh^2\lambda~e^{-2td}}
\end{eqnarray}
The integrals converge quickly except at $x=\pm d/2$, where they
diverge.  A plot of the energy density for several choices of $m$
is shown in Figure~\ref{fig:energy density 2}. 

\begin{figure}

\begin{center}
\BoxedEPSF{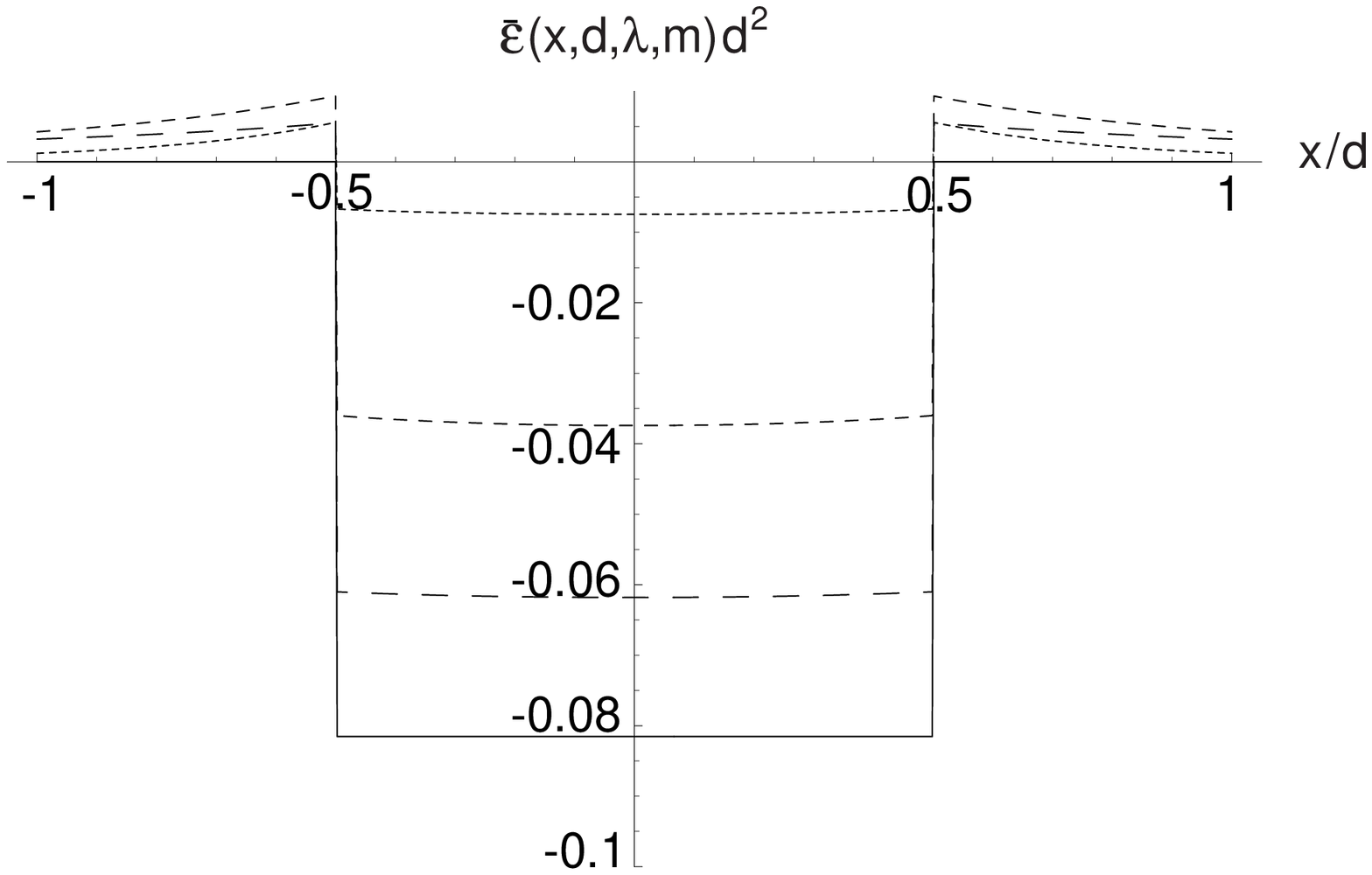 scaled 600}
\caption{For small $m$, the
interaction energy density is significant, and essentially
localized between the two spikes.  As $m$ grows larger, the
overall interaction term gets smaller, and energy density ``leaks
out'' to the region outside the spikes.  The plot shows $m=0$
(solid), $m=0.1/d$ (dashed), $m=0.3/d$ (finely dashed),
 and $m=1.0/d$ (dotted), with $\lambda=1$ fixed.}
\label{fig:interaction-density-m}
\BoxedEPSF{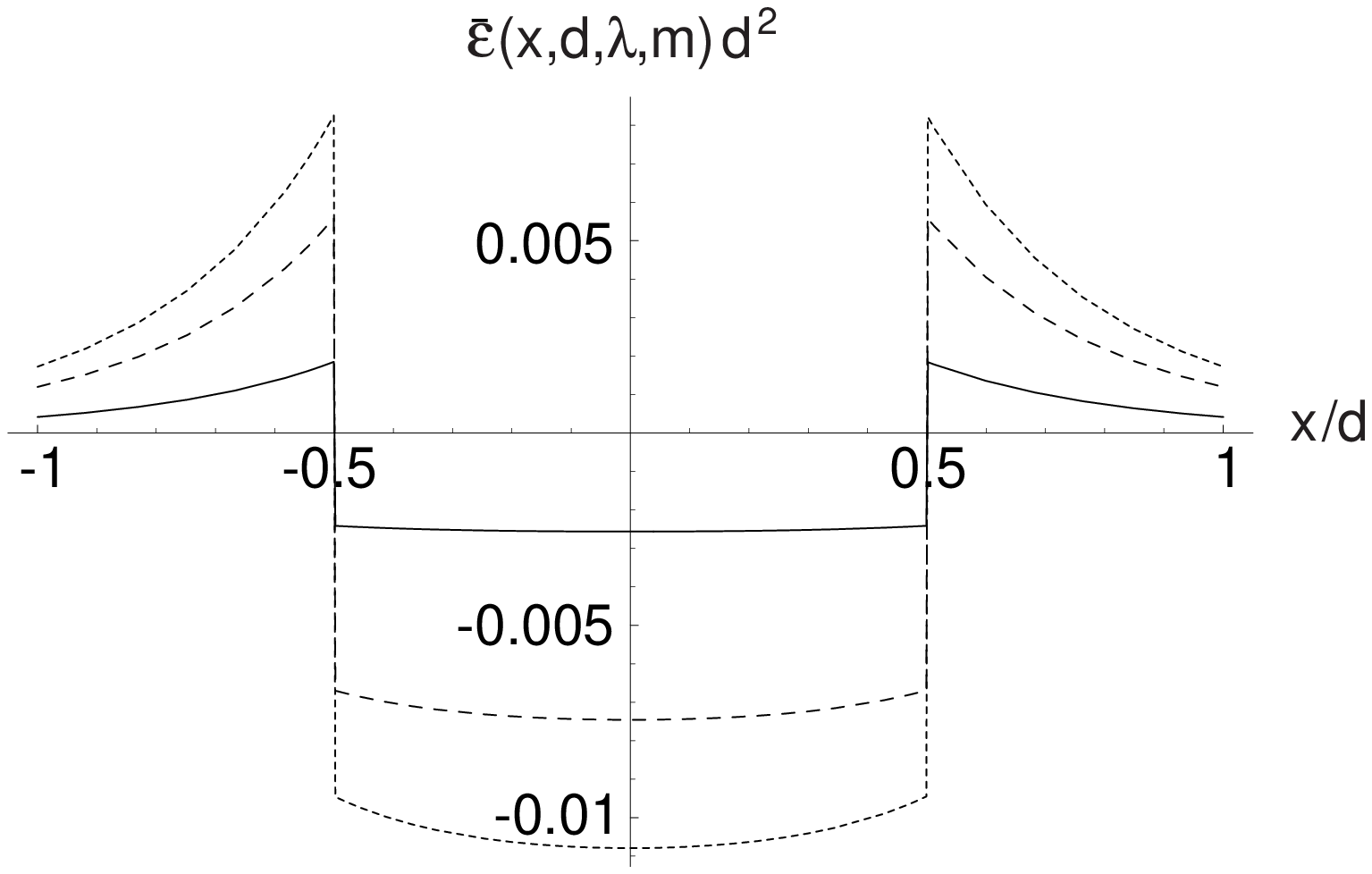 scaled 600}
\caption{Interaction energy density
as a function of $\lambda$.  The plot shows $\lambda=0.4$
(solid), $\lambda=1.0$ (dashed) and $\lambda\to\infty$ (dotted),
with $m=1$ fixed.  Note the discontinuities at $x=\pm d/2$.}
\label{fig:interaction-density-lambda}
\end{center}
\end{figure}

To eliminate local divergences which do not affect the force, we
consider the pure interaction energy density
$\bar{\epsilon}(x,d)$, defined as
\begin{equation}
\label{eq:bar-epsilon}
\bar{\epsilon}(x,d) =
\epsilon_2(x,d)-\epsilon_1(x-d/2)-\epsilon_1(x+d/2),
\end{equation}
where $\epsilon_1(x-d/2)$ is the energy density of a single spike
placed at $x=d/2$ defined in eq.~(\ref{eq:epsilon1}).  Because the
divergences are local to the points of interaction, this
subtraction will render the resulting expression finite
everywhere.  However, there is no reason to expect that the
energy density should be continuous across the point of
interaction, and this is indeed not the case.  In particular, for
$m=0$, the energy density vanishes in the region outside the
spikes, but takes a constant nonzero value in the region between
them.  Figures~\ref{fig:interaction-density-lambda}
and~\ref{fig:interaction-density-m}
show the interaction energy density in various configurations.

Finally we compute the force between two spikes.  Although the
total Casimir energy diverges, the divergences are localized to
the points of interaction and are identical to the divergences
associated with a single spike.  Therefore we define an
``interaction Casimir energy'' by subtracting away twice the
contribution from a single spike from the two spike Casimir
energy.  The subtracted quantity is independent of $d$ and
therefore does not contribute to the force.  To simplify the
computation we place the centers of the two subtracted spikes at
$x = \pm d/2$ respectively, and perform the subtraction within the
integrand of the expression for the energy.  As demonstrated in
Section~\ref{sc:energydensity}, this renders the energy density
finite everywhere.

We define the total interaction energy $E$ as the integral over
$x$ of $\bar{\epsilon}(x,d)$, defined by eq.~(\ref{eq:bar-epsilon}),
\begin{eqnarray}
\label{eq:total-interaction-energy}
E_{\rm int} & = & \int_{-\infty}^\infty dx\ \bar{\epsilon}(x,d) =
\frac{1}{\pi}\int_m^\infty dt\
2\sqrt{t^2-m^2}\sinh^2\lambda~e^{-2td} \times \nonumber \\
&\times& \frac{(m^2(td+1)-t^3d)\cosh\lambda-
(md(t^2-m^2)-mt)\sinh\lambda} {(t\cosh\lambda+m\sinh\lambda)^3+
(t^2-m^2)(t\cosh\lambda+m\sinh\lambda)e^{-2td}}
\end{eqnarray}
Since this integral is absolutely convergent, we can compute the
force $F(\lambda, m, d) = - {\rm d}E_{\rm int}/{\rm d}d$ by moving the
derivative operator under the integral sign.  The resulting
expression in not very enlightening, and the $t$-integration must
be performed numerically, however in certain interesting limits,
the formula simplifies greatly.
%

In the limit $\lambda \to \infty$, corresponding to impenetrable
spikes, or equivalently, bag boundary conditions, the total
interaction energy, eq.~(\ref{eq:total-interaction-energy})
reduces to reduces to
\begin{align}
        \lim_{\lambda\to\infty}E_{\rm int}(\lambda,m,d)
        &\equiv E_{\rm bag}(m,d)\\
        &= \frac{2}{\pi}\int_m^\infty
        \sqrt{t^2-m^2}\frac{m-d(t^2-m^2)}{t^2-m^2+(t+m)^2e^{2td}}dt
\end{align}
and the force simplifies as well,
\begin{align}
        \lim_{\lambda\to\infty}F(\lambda,m,d)
        &\equiv F_{\rm bag}(m,d)\\
        &= -\frac{2}{\pi}\int_m^\infty
        \sqrt{t^2-m^2}\frac{e^{2td}[(m^2-t^2)+2t^3d-2mt(md+1)]-(t-m)^2}
{[(t-m)+(t+m)e^{2td}]^2}dt
\end{align}

These integrals can be done analytically for $m=0$,
\begin{eqnarray}
E_{\rm bag}(0,d) &=& -\frac{\pi}{24d}\\
F_{\rm bag}(0,d) &=& -\frac{\pi}{24d^2},
\end{eqnarray}
an attractive inverse-square dependent force law.

\begin{figure}
\begin{center}
\BoxedEPSF{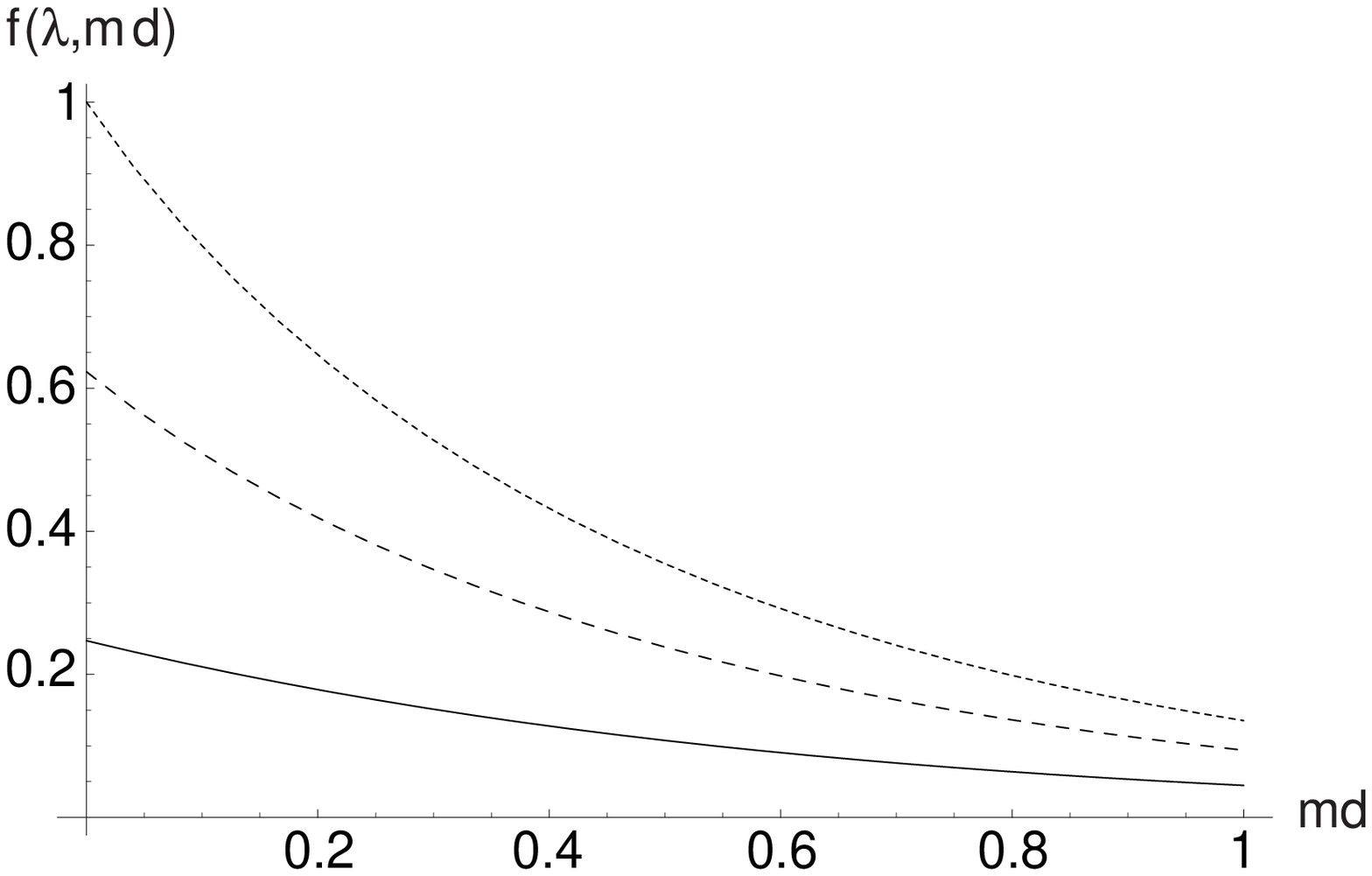 scaled 600}
\caption{Force as a function of separation between the spikes.
We write
$F = -\frac{\pi}{24d^2}f(\lambda,md)$, and the plot shows $f$ for
$\lambda=0.5$ (solid), $\lambda=1$ (dashed), and $\lambda \to
\infty$ (dotted). }
\label{fig:force-1}
\end{center}
\endfigure
\figure
\begin{center}
\BoxedEPSF{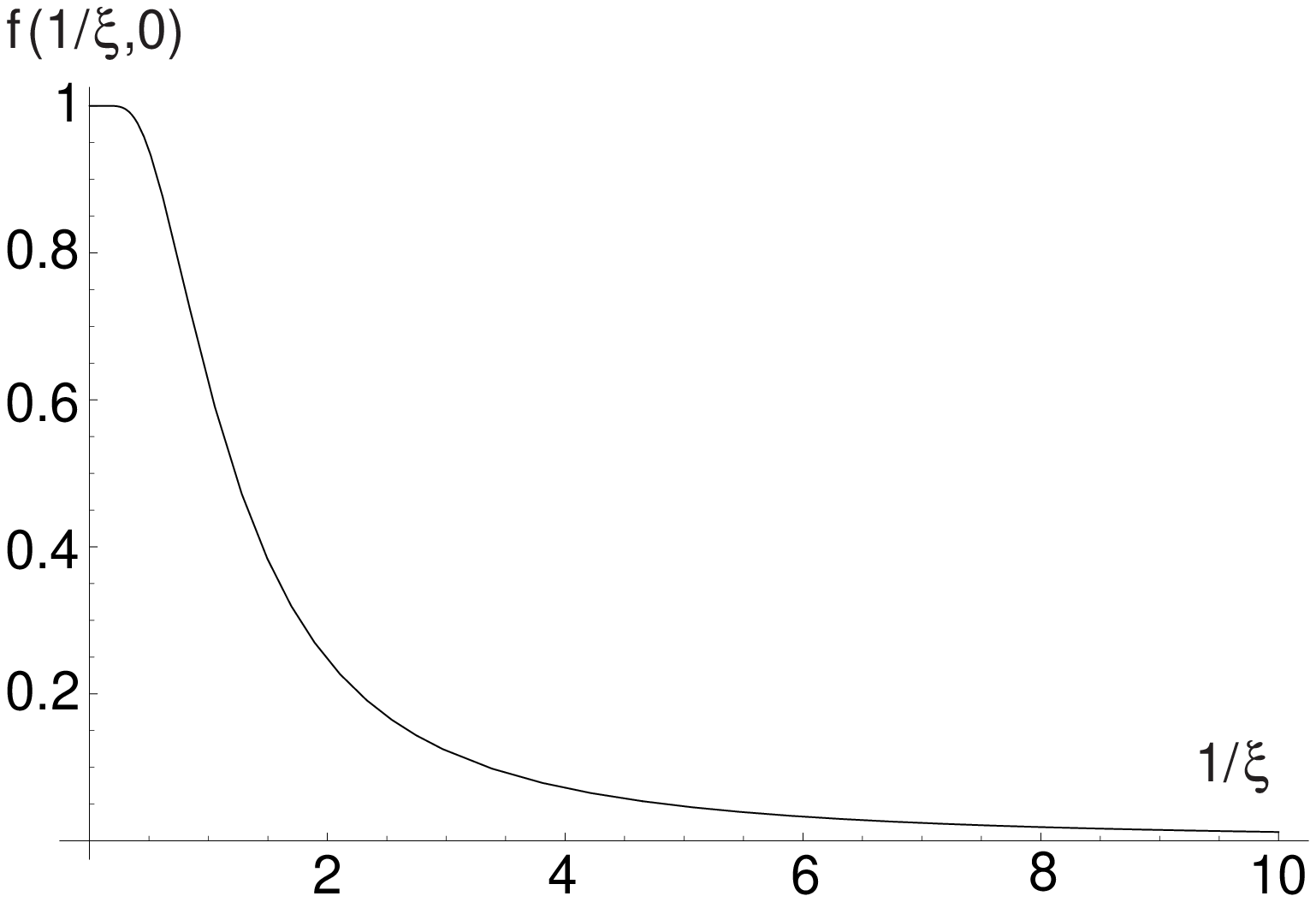 scaled 600}
\caption{For $m=0$, the force takes the form $F=-\frac{\pi}{24d^2}f(\lambda)$
for dimensional
reasons. This plot shows $f(1/\xi)$ as a function of $\xi =
1/\lambda$.}
\label{fig:force-2}
\end{center}
\end{figure}

In general, the force can be parameterized by
\begin{equation}
        F(\lambda, m ,d) = -\frac{\pi}{24d^2}f(\lambda, md),
\end{equation}
where $f$ is a dimensionless function of two dimensionless parameters
and describes the departure from the force in the massless case.
Figure~\ref{fig:force-1} shows a plot of $f(\lambda, md)$, and
Figure~\ref{fig:force-2} shows $f(\lambda,0)$.  $f(\lambda,
md)$ approaches a limit exponentially as $\lambda\to\infty$, $
f(\lambda, 0) \sim f(\infty, 0) + C~e^{-\lambda} $. As a consequence,
all derivatives of $f$ go to zero as $\lambda \to \infty$.  This is
visible in Figure~\ref{fig:force-2}.

\section{Discussion}

In this paper we have applied the methods of Ref.~\cite{longpaper}
to fermion fields in one dimension.  The background fields were
taken to be sharp spikes, imposing jump conditions on the dynamical
fermion field.  Using the method of computing the density of
states directly from the S matrix, we showed that the total
energy of such a configuration diverges and thus cannot be computed
directly using the jump condition approach.  The details of the
interaction cannot be ignored, and the divergences introduced by a sharp
potential in a fermion field are even worse than the ones that
appear when the dynamical field is a scalar.

Nonetheless, energy densities and forces are still well defined.
Using a formalism based on the Greens function, we showed how to
compute these quantities even when the background field is highly
singular, and derived expressions for the energy density in
backgrounds of one and two spikes.  The work culminated in a
computation of the force between two spikes in the presence of a
fermion field.  For a massless field and spikes implementing bag
boundary conditions, dimensional considerations require the force
to be an inverse square force as a function of the distance
between the spikes.  We explicitly verified this, and computed
the constant of proportionality.

\section{Acknowledgements}

We gratefully acknowledge discussions with V.~Khemani, N.~Graham,
M.~Quandt and E.~Farhi.  This work is supported in part by the
U.S.~Department of Energy (D.O.E.) under cooperative research
agreements~\#DF-FC02-94ER40818.

\end{document}